\documentclass[letterpaper, 10 pt, conference]{ieeeconf}  % Comment this line out if you need a4paper
\IEEEoverridecommandlockouts
\usepackage{cite}
\usepackage{amsmath,amssymb,amsfonts}
\usepackage{algorithmic}
\usepackage{graphicx}
\usepackage{subfigure}
\usepackage{textcomp}
\usepackage{booktabs}
\usepackage{epstopdf}
\usepackage{array}
\usepackage{algorithm}
\usepackage{pgffor}
\usepackage{bm}
\usepackage{multirow}
\usepackage{subcaption}
\usepackage{tablefootnote}
\usepackage{algorithm}

\usepackage[colorlinks,citecolor=green,urlcolor=blue,bookmarks=false,hypertexnames=true]{hyperref} 
\def\BibTeX{{\rm B\kern-.05em{\sc i\kern-.025em b}\kern-.08em
		T\kern-.1667em\lower.7ex\hbox{E}\kern-.125emX}}

\title{TV-Regularized Frequency-Domain Full-Waveform Inversion for Single-Sided Linear Ultrasound Array Data}
\author{Rui Guo, 
	Ditza Auerbach, and 
	Yonina C. Eldar.
	\thanks{This research was supported by the European Research Council (ERC) under the European Union’s Horizon 2020 research and innovation program (grant No. 101000967), by the Israel Science Foundation (grant No. 3805/21, 536/22) within the Israel Precision Medicine Partnership (IPMP) program, and by the Manya Igel Centre for Biomedical Engineering and Signal Processing.} 
	\thanks{Faculty of Math and Computer Science, Weizmann Institute of Science, Rehovot, Israel.}
}
\begin{document}

	\maketitle
	\begin{abstract}
		
		Quantitative speed-of-sound (SoS) and attenuation of tissues are closely related to pathology; however, conventional B-mode images are limited to qualitative visualization.
		Existing ultrasound full-waveform inversion (FWI) methods for quantitative SoS reconstruction  are primarily developed under  double-sided or ring-shaped arrays, which limits their applicability to  widely adopted routine clinical acquisitions.
		In this work, we develop a frequency-domain, total variation (TV)-regularized FWI framework tailored for single-sided linear ultrasound arrays, which enables quantitative reconstruction of SoS  maps using standard clinical probes.
		To address the severe ill-posedness and computational challenges in this setup, efficient forward modeling, fast gradient evaluation, ADMM-based optimization, and multi-GPU parallelization are integrated into the inversion framework.
		Numerical experiments in a thyroid cyst imaging scenario demonstrate that the proposed method  reconstructs the SoS of both simple (fluid-filled) and solid cysts with improved visual and quantitative performance compared to conventional FWI. Additional 2D and 3D simulations across different target and array apertures further elucidate the capabilities and limitations of single-sided ultrasound FWI.
	\end{abstract}
	
	\section{Introduction}
	% Replace below: The values of acoustic parameters, such as ...
	Acoustic parameters, such as speed-of-sound (SoS) and attenuation, are closely linked to pathological conditions, including fibrosis, tumors, and inflammation \cite{o2007ultrasound, mojabi2019experimental}. In clinical practice, these parameters are typically qualitatively inferred from 
	% changed below
	the B-mode ultrasound image echo intensity variation. However, explicit quantification in terms of the localized physical properties is lacking.  
	% till here
	In thyroid imaging, for instance, while simple cysts and solid nodules may be distinguished by their echo patterns, their apparent morphology and depth in B-mode images are often geometrically inaccurate due to the assumption of homogeneous SoS during beamforming. 
	To faciliate more precise differentiation of structures, lesion composition and complexity, 
	ultrasound computed tomography (USCT) has emerged as a powerful technique for quantitative SoS reconstruction \cite{guasch2020full, li2023quantitative,yan2025review}. In USCT, acoustic waves are transmitted through the domain of interest from multiple angles, and the recorded wavefields are used to reconstruct spatial maps of acoustic properties. Full-waveform inversion (FWI) has become a key reconstruction technique in USCT by iteratively matching measured and simulated waveforms through numerical solutions of the wave equation.
	
	Most current USCT systems rely on dense acquisition geometries with sufficient apertures, often involving hundreds or even thousands of transducer elements distributed around the imaging domain \cite{guasch2020full, zapf2022realization}, similar to those used in CT or MRI. Under such conditions, the abundance of multi-angle transmission data provides rich wavefield information, and reasonable reconstruction quality can often be achieved without strong regularization \cite{li2023quantitative}. However, such USCT data acquisition requires specialized hardware that is  incompatible with widely used clinical ultrasound systems operating in single-sided pulse–echo mode. Consequently, the translation of ultrasound FWI from USCT-oriented experimental platforms to routine clinical practice remains limited. 
	%split to 2 sentences
	In addition, USCT's reliance on transmission-mode signals makes it highly sensitive and often prohibitive to explore regions involving air–tissue interfaces. Examples are the trachea or digestive tract, whose strong acoustic impedance mismatches induce severe scattering and shadowing that exacerbate forward modeling errors and nonlinearities in FWI. 
	% In addition, the USCT data acquisition  requires specialized hardware that is  incompatible with popular clinical systems operating in single-sided pulse–echo mode. Consequently, the translation of ultrasound FWI from USCT-oriented experimental platforms to routine clinical practice remains limited.

	%Many FWI approaches for USCT rely on full-bandwidth, time-domain solvers based on finite-difference or difference-operator discretizations that were originally designed for strongly heterogeneous media. While effective in geophysical settings, such discretization strategies are not always strictly necessary for biomedical ultrasound imaging and may introduce avoidable computational overhead, thereby limiting real-time or time-efficient performance.

	SoS reconstruction for single-sided pulse–echo array data is attractive due to its compatibility with current clinical ultrasound systems \cite{shultzman2022nonlinear}. A variety of reconstruction strategies have been developed, including ray-based tomography     \cite{jaeger2015computed,sanabria2018spatial,stahli2020improved,jaeger2022pulse}, ultrasound refractor models \cite{jiang2024demonstration,martiartu2022toward,sanabria2018speed}, and more recently, data-driven deep learning approaches \cite{bernhardt2020training,sharon2024real}. Ray-based and reflector-based methods typically rely on high-frequency or straight-ray approximations of wave propagation, which significantly reduce computational complexity and enable real-time reconstruction. However, these approaches often neglect diffraction and refraction effects, which can lead to quantitative errors and limited spatial resolution especially in complex heterogeneous tissues.
	Deep learning–based inversion methods have also been proposed by learning mappings from ultrasound measurements to SoS distributions. While promising, the effectiveness of data-driven approaches remains closely linked to the availability of high-fidelity, physics-consistent training datasets and the interpretability of the solution process. Moreover, the severe ill-posedness induced by limited aperture and pulse–echo configurations  suggests that learning-based methods can significantly benefit from integration with physics-based frameworks  \cite{shlezinger2022model,guo2023physics}.
	
	In this context, FWI offers a physically rigorous backbone that complements existing ray-based and data-driven strategies. Directly employing existing FWI frameworks that have been advanced for USCT are insufficient for clinical pulse-echo arrays due to the following computational and physical challenges.
	Clinical ultrasound arrays typically operate at megahertz frequencies and target regions of interest that are deep under the skin, corresponding to imaging domains spanning hundreds or even thousands of wavelengths. Consequently, single-sided ultrasound FWI naturally constitutes a time and memory consuming inverse problem. Full-bandwidth, time-domain solvers widely adopted in seismic and USCT FWI \cite{ wang2025mask, cohen2025deep} work well for lower-frequency excitations and smaller investigated domains, but become computationally prohibitive when applied to clinical pulse–echo array data.
	In addition, existing clinical ultrasound probes typically provide limited aperture measurements, where the array length is often much shorter than the imaging depth. This limited-view geometry significantly increases the ill-posedness of single-sided ultrasound FWI and exacerbates non-uniqueness in deep target reconstruction \cite{wang2025mask}. These characteristics indicate that a tailored FWI framework for single-sided clinical ultrasound imaging is necessary.
	
	In this work, we present a frequency-domain FWI framework that enables quantitative SoS reconstruction from single-sided clinical ultrasound array data. By exploiting the relatively homogeneous background SoS in soft tissues, our approach employs an integral formulation in the forward model, which leverages translational invariance of the Green’s function to accelerate the computation process. To reduce the ill-posedness in the inverse problem, we formulate a total variation (TV) constrained optimization problem solved using the alternating direction method of multipliers (ADMM). Computational efficiency is further enhanced through a hybrid approach for Jacobian matrix computation, implicit schemes for Hessian and gradient computation, and multi-GPU parallelization.
	
	Our contributions  are summarized as follows:
	\begin{itemize} 
		\item A systematic investigation of pulse–echo FWI in the context of single-sided arrays—a highly ill-posed but clinically essential scenario previously under-explored.
		
		\item An integrated, robust inversion pipeline that integrates frequency-domain modeling with sparsity-promoting regularization and efficient computation techniques.
		
		\item Evaluation of the proposed framework in 2D and 3D simulations across different targets and array apertures	to offer insights into the capabilities and boundaries of single-sided ultrasound FWI.
	\end{itemize}

	The paper is organized as follows. Section~\ref{SecII} introduces the measurement setup and forward model. Section~\ref{SecIII} formulates the inverse problem and presents the ADMM-based optimization algorithm. Section~\ref{SecIV} presents several fast computation techniques.  Section~\ref{SecV} presents numerical validations and performance evaluations under different imaging conditions. Finally, Section~\ref{SecVI} concludes the paper and discusses potential future work.

	\section{Problem statement}\label{SecII}
	
	%	The forward model enables to simulate the ultrasound measurements for successive transmissions in a given medium, which builds the foundations for FWI.
	
	We consider acoustic wave propagation in an inhomogeneous domain $V\subset \mathbb R^{d}$ ($d\in \{2,3\}$ corresponding to 2D or 3D domains) with constant density $\rho_0$ ($kg/m^3$) but spatial-varying SoS $c(\mathbf x)$ ($m/s$) and attenuation coefficient $\alpha (\mathbf {x})$ ($Np/m/Hz$), where $\mathbf x\in V$ denotes the spatial locations within the region enclosing all inhomogeneities. The medium can be characterized by complex  compressibility defined by
	\begin{equation}
		\kappa(\mathbf x) = \kappa'(\mathbf x) - \mathrm{j}\,\kappa''(\mathbf x) ,
	\end{equation}
	with
	\begin{equation}\label{eq:eqsos}
		\kappa'(\mathbf{x})= \frac{1}{c(\mathbf{x})^2 \rho_0} - \frac{\alpha(\mathbf{x})^2}{(2\pi)^2 \rho_0}
		\approx \frac{1}{c(\mathbf{x})^2 \rho_0}, 
		\quad
		\kappa''(\mathbf{x}) = \frac{ \alpha(\mathbf{x})}{\pi c(\mathbf{x}) \rho_0},
	\end{equation}
	We decompose the inhomogeneous $\kappa(\mathbf x)$ into a homogeneous background and a scattering perturbation \cite{mojabi2015ultrasound,song2020study}.  The former is characterized by  the background compressibility $\kappa_0$ related to  a constant background SoS $c_0$, an attenuation coefficient $\alpha_0$, and density $\rho_0$, while the latter is described by the contrast function\cite{song2020study},
	\begin{equation}\label{key}
		m(\mathbf{x}) = {\kappa(\mathbf{x})}/{\kappa_0} - 1.
	\end{equation}
	
	The frequency-domain scattered field for the $i$-th transmission, $j$-th receiver, at the $l$-th frequency point, with  $i = 1,\ldots,N_T$, $j = 1,\ldots,N_R$, and  $l = 1,\ldots,N_F$, is modeled by \cite{wiskin20173,song2020study,zhou2023frequency}
	\begin{equation}\label{eq:data_eq}
		P^{\mathrm{s}}_{i,l,j}
		= k_{l, 0}^2
		\int_{x\in \Omega_j}
		\int_{x'\in V}
		G_{l,0}(\mathbf{x}, \mathbf{x}')\,
		m(\mathbf{x}')\, p_{i}(\mathbf{x}', \omega_l)\,
		d\mathbf{x}'\, d\mathbf{x},
	\end{equation}
	where $\omega_l$ denotes the $l$-th angular frequency, $k_{l,0}^2 = \omega_l^2 \kappa_0 \rho_0$, domain $\Omega_j$ denotes the sensing region occupied by the $j$th receiver located on part of the edge of domain $V$, and $G_{l,0}(\mathbf{x}, \mathbf{x}')$ denotes the Green’s function related to  $k_{l,0}$ (which is analytically known). $p_i(\mathbf{x}, \omega_l)$ is called total pressure field that satisfies 
	\begin{equation}
		p_i(\mathbf{x},\omega_l)
		= p_{i,0}(\mathbf{x},\omega_l)
		+ k_{l,0}^2 \int_V
		G_{l,0}(\mathbf{x}, \mathbf{x}')\,
		m(\mathbf{x}')\, p_i(\mathbf{x}',\omega_l)\,
		d\mathbf{x}',
		\label{eq:vie}
	\end{equation}
	where $p_{i,0}(\mathbf{x},\omega_l)$ denotes the incident field due to the $i$-th transmission in the homogeneous background characterized by $\kappa_0$. 
	
	Equations~\eqref{eq:data_eq} and \eqref{eq:vie} offer computational advantages over  time-domain differential-equation-based modeling.
	These formulations perform discrete frequency sampling  instead of full-bandwidth time-domain propagation, allow discretization only the inhomogeneous region to reduce memory usage, and can leverage the translational invariance of the Green’s function to accelerate convolutional operators via Fast Fourier Transform (FFT) (details in Section~\ref{SecIV}).
	
	%	The pressure field $p_i(\mathbf{x},\omega_l)$ for all $\mathbf x\in V$ can be numerically solved from a  matrix equation established according to \eqref{eq:vie}. Then, the scattered field caused by medium inhomogeneity, which is detected  at the $j$th receiver ($j=1,\ldots,N_R$), can be modeled  by \cite{song2020study}
	
	We aim to recover the contrast distribution $m(\mathbf x)$, $\mathbf x \in V$  from all frequency-domain measurements $P^{\mathrm{s}}_{i,l,j}$ acquired by a single-sided ultrasound array, after which  SoS can be infered from \eqref{eq:eqsos}. In practice, only time-domain channel measurements for multiple transmit--receive pairs are recorded. Frequency-domain measurements at selected angular frequencies $\{\omega_l\}_{l=1}^{N_F}$ can be obtained via Fourier transformation from the recorded time-domain channel data.
	
	For notational convenience, all frequency-domain measurements are stacked into a complex-valued data vector
	$\mathbf d_{\mathrm{obs}} = \{ P^{\mathrm s}_{i,j,l} \}_{i,j,l} \in \mathbb C^{N_T N_R N_F}$, and the unknown contrast distribution is discretized as a vector $\mathbf m \in \mathbb C^{N_e}$, where $N_e$ denotes the number of pixels ($d=2$) or voxels ($d=3$) in the imaging domain $V$. We define a nonlinear forward operator $F(\cdot)$ that maps $\mathbf m$ to all frequency-domain measurements  according to \eqref{eq:data_eq} and \eqref{eq:vie}. The forward model can then be written compactly as
	\begin{equation}\label{key}
		\mathbf d_{\mathrm{obs}} = F(\mathbf m) + \mathbf n,
	\end{equation}
	where $\mathbf n$ denotes additive Gaussian noise. In the following, we focus on estimating $\mathbf m$ from the frequency-domain data $\mathbf d_{\mathrm{obs}}$.
	
	\section{Full-wave inversion algorithm}\label{SecIII}
	We formulate a regularized optimization problem that
	includes data fidelity and  prior information:
	\begin{equation}\label{eq:costfunction1}
		\begin{aligned}
			\min_{\mathbf m, \mathbf g}\;
			& \frac{1}{2}\|{F}(\mathbf m)-\mathbf d_{\mathrm{obs}}\|_2^2
			+ \lambda \|\mathbf g\|_{1,2}
			+ \frac{\gamma}{2}\|\mathbf m-\mathbf m_{\mathrm{ref}}\|_2^2,\\
			\quad
			&\text{s.t. } \mathbf g = \nabla \mathbf m,
		\end{aligned}
	\end{equation}
	where the first term enforces consistency between the simulated and observed
	scattered fields, the second term promotes piecewise-smoothness
	through  isotropic TV \cite{lou2015weighted}, and the third term incorporates a prior reference model $\mathbf m_{\mathrm{ref}}$ ($\gamma=0$ if no reference is available).
	
	Here, the discrete gradient operator $\nabla$ maps $\mathbf{m}$ to the spatial gradient in $d$ dimensions,  i.e., $\nabla \mathbf m = [\partial_1 \mathbf m;\cdots;\partial_d \mathbf m]\in \mathbb C^{N_e\times d}$. To handle the sparsity constraint, an auxiliary variable $\mathbf g = [\mathbf g_1, \cdots, \mathbf g_d]\in \mathbb C^{N_e \times d}$ is introduced. The isotropic TV is achieved via the $\ell_{1,2}$ norm defined as  $\|\mathbf g\|_{1,2}
	=
	\sum_{n=1}^{N_e}
	\left(
	\sum_{k=1}^{d} |g_{n,k}|^2
	\right)^{1/2}$.  The
	regularization parameters $\lambda$ and $\gamma$ control the relative strength
	of the TV prior and the reference-model constraint, respectively.
	
	The augmented Lagrangian for solving \eqref{eq:costfunction1} is given by
	\begin{equation}
		\begin{aligned}
			\mathcal{L}(\mathbf m, \mathbf g, \mathbf u)
			=
			& \frac{1}{2}\|{F}(\mathbf m) - \mathbf d_{\mathrm{obs}}\|_2^2
			+ \lambda \|\mathbf g\|_{1,2}
			\\ &+ \frac{\gamma}{2}\|\mathbf m - \mathbf m_{\mathrm{ref}}\|_2^2
			+ \frac{\rho}{2}\|\mathbf g - \nabla \mathbf m + \mathbf u\|_2^2,
		\end{aligned}
	\end{equation}
	where  $\rho>0$ is the penalty parameter controlling the enforcement of the
	constraint, and $\mathbf u$ denotes the scaled dual variable.
	It is minimized using ADMM in an iterative manner \cite{neal2011distributed}.  At iteration $k$, the algorithm updates the variables sequentially, denoted as $\mathbf m_k$, $\mathbf g_k$, and $\mathbf u_k$, respectively.  
	
	\textbf{Update  $\mathbf m$.}  
	The contrast $ \mathbf m$ is updated by solving the nonlinear optimization problem
	\begin{equation}\label{eq:m-step}
		\begin{aligned}
			\mathbf m_{k+1}
			=&
			\arg\min_{\mathbf m}
			\;
			\frac{1}{2}\|{F}(\mathbf m) - \mathbf d_{\mathrm{obs}}\|_2^2\\& 
			+ \frac{\gamma}{2}\|\mathbf m - \mathbf m_{\mathrm{ref}}\|_2^2 
			+ \frac{\rho}{2}\|\mathbf g_k - \nabla \mathbf m + \mathbf u_k\|_2^2.
		\end{aligned}
	\end{equation}
	To solve this subproblem, we employ a Gauss--Newton strategy by linearizing the nonlinear forward operator $	{F}(\mathbf m)
	\approx  {F}(\mathbf m_k) + \mathbf J_k (\mathbf m - \mathbf m_k)$,	where $\mathbf J_k$ denotes the Jacobian (Fr\'echet derivative) of
	$ {F}(\cdot)$ evaluated at $\mathbf m_k$, whose column is given by $	{\delta P_{i,j,l}^{\mathrm{s}}}/{\delta \mathbf m_k}$. 
	Substituting this linearization into~\eqref{eq:m-step} 	yields 
	\begin{equation}\label{key}
		\mathbf m_{k+1} = \mathbf m_k + \Delta\mathbf m,
	\end{equation}
	where  the update is solved by the following normal equation 
	\begin{equation}\label{eq:normal_eq0}
		\mathbf A_k \, \Delta\mathbf m = \mathbf b_k,
	\end{equation}
	with
	\begin{equation}\label{eq:normal_eq}
		\begin{aligned}
			\mathbf A_k
			&=
			\mathbf J_k^{H}\mathbf J_k
			+ \rho \nabla^T\nabla
			+ \gamma \mathbf I, \\[4pt]
			\mathbf b_k
			&=
			\mathbf J_k^{H}\bigl(\mathbf d_{\mathrm{obs}} - {F}(\mathbf m_k)\bigr)
			+ \rho \nabla^T\bigl(\mathbf g_k + \mathbf u_k - \nabla \mathbf m_k\bigr)\\&
			+ \gamma \bigl(\mathbf m_{\mathrm{ref}} - \mathbf m_k\bigr),
		\end{aligned}
	\end{equation}
	where  $\nabla^T$ operator is defined as
	$\nabla^T \mathbf a = \sum_{q=1}^{d} \partial_q^{\,T}\mathbf a_q$ for arbitary $\mathbf a = [\mathbf a_1;\ldots;\mathbf a_d]\in \mathbb C^{N_e\times d} $. 
	
	By incorporating second-order information through the Jacobian, the Gauss-Newton approach achieves faster convergence than first-order or quasi-Newton schemes such as in  \cite{wiskin20173,zhou2023frequency, song2020study, li2023quantitative}. This reduces the number of iterations and the computational burden of full forward simulations, which is crucial in large-scale array modeling \cite{yan2025review}. Efficient Jacobian computation and the solution of the associated normal equation  are detailed in Section~\ref{SecIV}.
	
	\textbf{Update   $\mathbf g$.}  
	Given the updated contrast $\mathbf m_{k+1}$, the auxiliary variable is
	obtained by solving
	\[
	\mathbf g_{k+1}
	=
	\arg\min_{\mathbf g}
	\;
	\lambda \|\mathbf g\|_{1,2}
	+ \frac{\rho}{2}\| \mathbf g - \nabla \mathbf m_{k+1}  + \mathbf u_k\|_2^2.
	\]
	This step promotes joint sparsity across spatial dimensions and admits a closed-form solution via the isotropic shrinkage operator:
	\begin{equation}
		\mathbf{g}_{k+1} = \operatorname{shrink}_{\frac{\lambda}{\rho}} \left( \nabla \mathbf{m}_{k+1} - \mathbf{u}_k \right),
	\end{equation}
	where for each spatial element $n$ with local gradient vector $\mathbf{a}_n \in \mathbb{C}^d$, the operator is defined as:
	\begin{equation}
		\operatorname{shrink}_{\tau}(\mathbf{a}_n) = \max \left( 1 - \frac{\tau}{\|\mathbf{a}_n\|_2}, 0 \right) \mathbf{a}_n.
	\end{equation}
	
	\textbf{Update   $\mathbf u$.}  
	The scaled dual variable is updated by
	\[
	\mathbf u_{k+1}
	=
	\mathbf u_k
	+ \rho \big( \mathbf g_{k+1} - \nabla \mathbf m_{k+1} \big).
	\]
	
	The iterative process terminates when the data fidelity reaches a predefined threshold or the maximum iteration step is attained. Algorithm~\ref{alg:fwi} summarizes the proposed frequency-domain FWI procedure.
	
	\begin{algorithm}[t]
		\caption{Frequency-domain FWI with TV regularization}
		\label{alg:fwi}
		\begin{algorithmic}[1]
			\REQUIRE Frequency domain measurement $\mathbf d_{\mathrm{obs}}$, initial model $\mathbf m_0$, reference model $\mathbf m_{\mathrm{ref}}$ (optional)
			\REQUIRE Regularization parameters $\lambda, \rho, \gamma$
			\STATE Initialize $\mathbf m_0$, $\mathbf g_0 = \nabla \mathbf m_0$, $\mathbf u_0 = \mathbf 0$
			\FOR{$k = 0,1,\dots,K-1$}
			\STATE \textbf{Forward simulation:} compute ${F}(\mathbf m_k)$
			\STATE \textbf{Model update:} solve \eqref{eq:normal_eq0} for $\Delta \mathbf m$, which requires \eqref{eq:sensitivity}\eqref{eq:J_forward}\eqref{eq:J_adjoint}
			\STATE \textbf{Update $\mathbf m$:} $\mathbf m_{k+1} = \mathbf m_k + \Delta \mathbf m$
			\STATE \textbf{Update $\mathbf g$:} $\mathbf g_{k+1} \leftarrow \operatorname{shrink}_{\lambda/\rho}(\nabla \mathbf m_{k+1} - \mathbf u_k)$
			\STATE \textbf{Update $\mathbf u$:} $\mathbf u_{k+1} = \mathbf u_k + \rho(\mathbf g_{k+1} - \nabla \mathbf m_{k+1})$
			\ENDFOR
			%			\STATE 
			\RETURN reconstructed contrast $\mathbf m_K$
		\end{algorithmic}
	\end{algorithm}

	\section{Efficient computation techniques}\label{SecIV}
	This section introduces several techniques that overcome the intensive memory requirement and  computing time to implement our algorithm.
	
	{The discretization of $m(\mathbf x)$, $p_i(\mathbf x, \omega_l)$ and  $p_{i,0}(\mathbf x, \omega_l)$ used for solving \eqref{eq:vie} and \eqref{eq:data_eq} is typically finer than the inversion mesh, since forward simulations must satisfy numerical stability requirements, whereas the medium contrast for inversion can often be represented on a coarser grid. In this section, with some abuse of notation, we use the continuous form of a variable, e.g., $m(\mathbf x)$, to denote discrete variables defined on the forward mesh, and the corresponding vector form, e.g., $\mathbf m \in \mathbb C^{N_e}$, to denote their representations on the inversion mesh.
	}. 
	
	\subsection{FFT accelerated forward solver }
	To efficiently compute the total field $p_{i}(\mathbf{x},\omega_l)$ from
	\eqref{eq:vie}, we employ an FFT-based acceleration of the volume integral.
	After discretization, \eqref{eq:vie} leads to a large linear system involving a
	dense Green's matrix, whose explicit formation and direct application would
	incur prohibitive computational and memory costs. By exploiting the translation
	invariance of the Green's function in a homogeneous background, the product of
	the Green's matrix and a vector is implemented as a convolution and evaluated
	efficiently using FFTs \cite{mojabi2015ultrasound,song2020study}. The resulting linear system is then solved using an
	iterative Krylov subspace method, such as GMRES, which requires only repeated
	matrix--vector products. This approach reduces the per-iteration computational
	complexity from $\mathcal O(N^2)$ to $\mathcal O(N\log N)$, where $N$ denotes the
	number of discretization points in the forward computational domain, and eliminates the
	need to explicitly store the dense Green's matrix.
	\subsection{Hybrid Born--Distorted Born Jacobian approximation}
	In the $k$-th iteration, the entries of $\mathbf J_k$ are computed using an adjoint-state method \cite{abubakar20082}. The derivative of the measurement
	$P^{\mathrm s}_{i,j,l}$ with respect to the $n$-th element of $\mathbf m_{k}$, denoted by $ m_{n}$, is
	approximated as
	\begin{equation}\label{eq:sensitivity}
		\frac{\delta P_{i,j,l}^{\mathrm{s}}}{\delta  m_{n}}
		\approx
		k_{0,l}^2
		\int_{\mathbf x'\in\tau_n}
		p_i(\mathbf x',\omega_l)\,
		\tilde{p}_j(\mathbf x',\omega_l)\,
		d\mathbf x',
	\end{equation}
	where $p_i(\mathbf x',\omega_l)$ and $\tilde{p}_j(\mathbf x',\omega_l)$ respectively denotes the pressure field generated by the $i$-th transmision and the field of
	an adjoint source virtually placed at receiver $j$ \cite{abubakar20082}, computed under the iterate $\mathbf m_k$. Upon discretization, each Jacobian
	entry corresponds to the sensitivity with respect to a pixel/voxel,
	and the above integral is evaluated over the associated discretization cell
	$\tau_n$.
	The derivative corresponds to a distorted  Born
	linearization of the forward model \cite{hesford2010fast}. 
	
	A strict computation of
	$\mathbf J_k$ would require both the forward fields
	$p_i(\mathbf x,\omega_l)$ and the adjoint fields
	$\tilde p_j(\mathbf x,\omega_l)$ at each iteration, which results in
	$(N_T+N_R)N_F$ forward simulations per iteration. To reduce this computational
	cost, we exploit the weak-scattering assumption and update $\tilde p_j(\mathbf x,\omega_l)$ 
	only every several iterations to correct
	nonlinear effects. Between these updates, the adjoint fields
	$\tilde p_j(\mathbf x,\omega_l)$ are kept fixed, which corresponds to a Born approximation for the adjoint problem.

	\subsection{Implicit Jacobian operations}
	The explicit assembly, factorization, or inverse  of  $\mathbf A_k$ is computationally prohibitive, which makes a direct solution of \eqref{eq:normal_eq0} infeasible. We therefore solve the linear system using an iterative
	Krylov method (e.g., GMRES or conjugate gradient), which requires only matrix--vector
	products. The $\mathbf J_k$ is not formed by \eqref{eq:sensitivity} explicitly; instead, we
	compute the matrix--vector product directly from $p_i(\mathbf x,\omega_l)$ 
	and $\tilde{p}_j(\mathbf x,\omega_l)$ as the following, with details presented in \cite{li2011compressed,guo2023three}.
	
	Given an contrast-space vector $\mathbf a\in\mathbb C^{N_e}$ and its corresponding representation on forward meth  $a(\mathbf x)$, the forward Jacobian
	product corresponding to the measurement indexed by $(i,j,l)$ is computed as
	\begin{equation}\label{eq:J_forward}
		(\mathbf J \mathbf a)_{i,j,l}
		=
		k_{0,l}^2 \int_{\mathbf x\in V}
		p_i(\mathbf x,\omega_l)\,
		\tilde{p}_j(\mathbf x,\omega_l)\,
		a(\mathbf x)\,
		d\mathbf x.
	\end{equation}
	
	Similarly, for a data-space vector $\mathbf b\in\mathbb C^{N_TN_RN_F}$ whose element is $b_{i,j,l}$, the adjoint Jacobian product for the $n$-th element is evaluated by
	\begin{equation}\label{eq:J_adjoint}
		(\mathbf J^{H} \mathbf b)_n
		=
		\int_{\mathbf x\in \tau_n} k_{0,l}^2 \sum_{i,j,l}
		\overline	{p_i(\mathbf x,\omega_l)}\,
		\overline{	\tilde{p}_j(\mathbf x,\omega_l)}\, b_{i,j,l}\, d\mathbf x.
	\end{equation}
	Additionally, $\mathbf J^{H}\mathbf J\mathbf a$ is obtained
	implicitly by successive multiplication, i.e.,
	$
	\mathbf J^{H}\mathbf J\mathbf a \;=\; \mathbf J^{H}(\mathbf J\mathbf a).
	$
	\subsection{Parallelization strategy}
	We adopt a source-based parallelization strategy. The $N_T$ transmitters are evenly
	distributed across different computing units, such that each worker
	handles an independent subset of sources. The corresponding primary fields
	$p_i(\mathbf x,\omega_l)$ are therefore partitioned across GPUs.
	The adjoint fields $\tilde{p}_j(\mathbf x,\omega_l)$ associated with all receivers
	are shared across all GPUs, as they are required for forming Jacobian-related
	matrix--vector products. During each iteration, partial contributions to the
	Jacobian operations are computed locally on each worker and subsequently
	aggregated to assemble the final matrix--vector product.  
	\section{Numerical examples}\label{SecV}
	We consider a simplified thyroid cyst imaging scenario, which is motivated by its clinical value in distinguishing between different cystic pathologies using quantitative acoustic parameters. Both 2D and 3D experiments are conducted to evaluate the performance  under different imaging conditions.
	
	In the 2D experiment, we first reconstruct a simple cyst and a solid cyst with a radius of 1~cm, where the former is fluid-filled and the latter contains a solid component. The two targets exhibit similar structural appearances but distinct acoustic properties. The study demonstrates our method's performance against conventional FWI without regularization and evaluates whether the internal SoS can reliably inverted between simple and solid cyst types. We then move to a combined muscle-cyst scenario to test the inversion performance for inhomogeneous background. 
	
	In the 3D experiment, we further assess the robustness of the proposed method under an extremely limited observation aperture and target radius less than $1 $ mm. This experiment aims to investigate the feasibility of the proposed approach to distinguish between two early-stage cystic targets in a clinically realistic and severely ill-posed setting.
	
	\subsection{2D scenario}
	\textbf{Measurement setup}: The measurements are simulated in the frequency domain using a set of 15 discrete frequencies uniformly distributed between 0.2~MHz and 3~MHz. The  linear array consists of 64 elements with an element spacing of 0.78~mm. Plane-wave transmissions are synthesized using the elements with steering angles uniformly spanning from $-60^\circ$ to $60^\circ$ with a total of 60 transmit events. All elements are used as receivers for each transmission. 
	
	\textbf{Single cyst scenario:} The simple cyst is modeled as a fluid-filled region with a uniform SoS of 1540~m/s and low acoustic attenuation of 0.02~dB/cm/MHz, representing homogeneous cystic fluid.
	The solid cyst is modeled as a region containing randomly distributed high-SoS scatterers, with an average SoS of 1620~m/s and an attenuation of 1.2~dB/cm/MHz, reflecting the presence of solid or highly viscous material within the cyst.
	The cyst wall is represented by a thin surrounding layer with a thickness of 0.8~mm and a SoS of 1580~m/s.
	The background medium is modeled with speckle-like scatterers to mimic tissue heterogeneity, while its average SoS is set to 1540~m/s and attenuation is set to 0.5 dB/cm/MHz.
	%	These two models are presented in Fig.~\ref{cyst_gt}(a) and Fig.~\ref{cyst_gt}(b), respectively.
	
	\begin{figure}[!t]
		%\centering
		%		\subfigure[]{	\includegraphics[trim=10 2 10 5, clip, width=0.9\linewidth]{simple_cyst_gt.png}	}
		%		\subfigure[]{	\includegraphics[trim=10 2 10 4, clip, width=0.9\linewidth]{solid_cyst_gt.png}	}
		{	\includegraphics[trim=10 2 2 6, clip, width=1\linewidth]{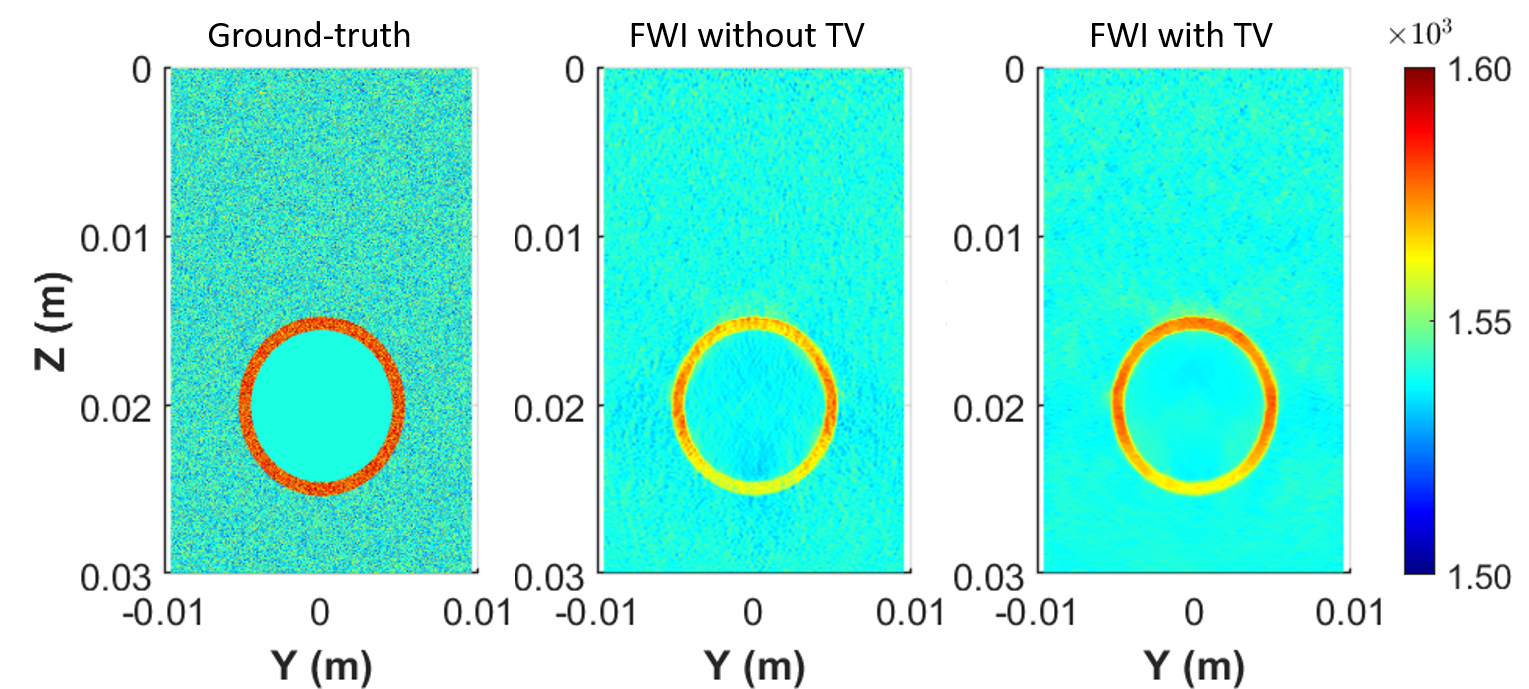}	}
		
		%    \vspace{-0.5cm}/
		\caption{Ground-truth and inverted SoS for the simple cyst scenario. Z-axis: depth.  Left: Ground-truth. Middle: inverted SoS from  FWI without TV regularization. Right: inverted SoS from FWI with TV regularization.   }
		\label{simple_cyst_compare} 
		%	\vspace{-0.4cm}
	\end{figure}

	The simulated data for for inversion are corrupted with 2\% Gaussian noise. We conduct FWI without or with TV regularization. Both approaches are run for a maximum of \( 20\) iterations, starting from the same homogeneous initial model with an SoS of 1540~m/s and zero attenuation. The reference model \(\mathbf{m}_{\mathrm{ref}}\) is set equal to the initial model, with a small regularization strength \(\gamma = 10^{-5}\).

	\begin{figure}[!t]
		%\centering
		%		\subfigure[]{	\includegraphics[trim=10 2 10 11, clip, width=0.9\linewidth]{simple_cyst_compare.png}	}
		%		\subfigure[]{	\includegraphics[trim=10 2 10 8, clip, width=0.9\linewidth]{simple_cyst_fwi.png}	}
		{	\includegraphics[trim=10 2 2 6, clip, width=1\linewidth]{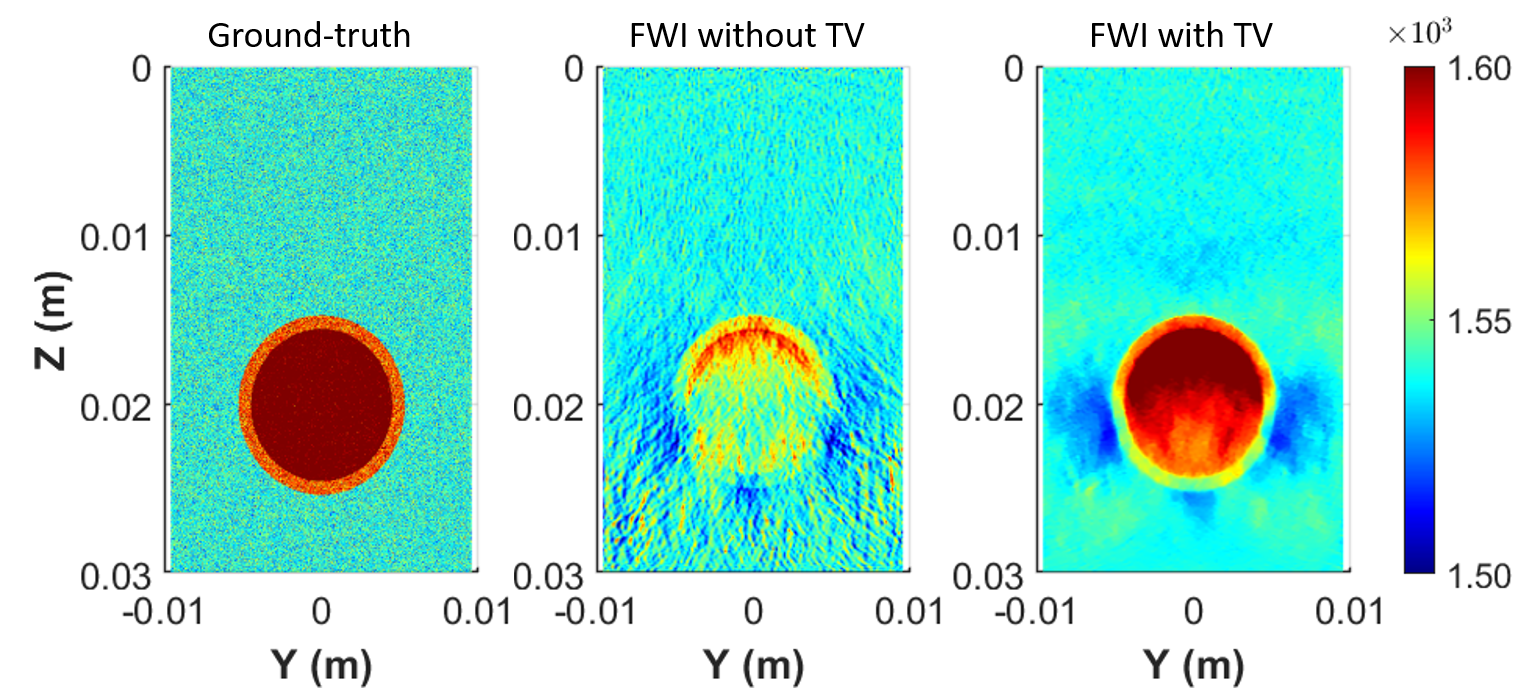}	}
		
		\caption{Ground-truth and inverted SoS for the solid cyst scenario.  Z-axis: depth. Left: Ground-truth. Middle: inverted SoS from  FWI without TV regularization. Right: inverted SoS from FWI with TV regularization. }
		\label{solid_cyst_compare} 
		%	\vspace{-0.4cm}
	\end{figure}
	%	\footnotetext{The imaginary part of the acoustic contrast is several orders of magnitude smaller than its real part, which limits the quantitative accuracy of attenuation reconstruction.}
	
	Figure~\ref{simple_cyst_compare} and Fig.~\ref{solid_cyst_compare} present the reconstruction results for the simple and solid cyst scenario, respectively. For the simple cyst case, FWI with TV successfully recovers a homogeneous speed-of-sound distribution within the cyst wall, consistent with the ground truth. 
	In contrast, the FWI without TV  exhibits artifacts inside the cystic region, indicating instability caused by the noise and the ill-posed nature of single-sided inversion. For the second case, due to the high SoS contrast of the solid cyst, acoustic waves have strong refraction and reflection at the cyst boundary, which significantly limits wave penetration into deeper regions and increase the ill-posedness of the problem. TV regularization stabilizes the solution process and recovers the high SoS region inside the solid cyst wall, while the basic FWI has a blurred and incomplete reconstruction of the solid inclusion. 	
	%	We also note that the attenuati
	%	Nevertheless, the normalized low-attenuation region reconstructed by the proposed method in Fig.~\ref{cyst_simple}(a) remains spatially consistent with the cyst location.
	
	During our experiments, we also observed that attenuation reconstruction is more challenging than the SoS reconstruction, which is likely because the imaginary part of the acoustic contrast is several orders of magnitude smaller than the real part and is therefore dominated by numerical errors \cite{song2020study}. We will address a more accurate inversion for the attenuation in a future work. 
	
	\begin{figure}[!t]
		%\centering
		{	\includegraphics[trim=10 2 3 5, clip, width=1\linewidth]{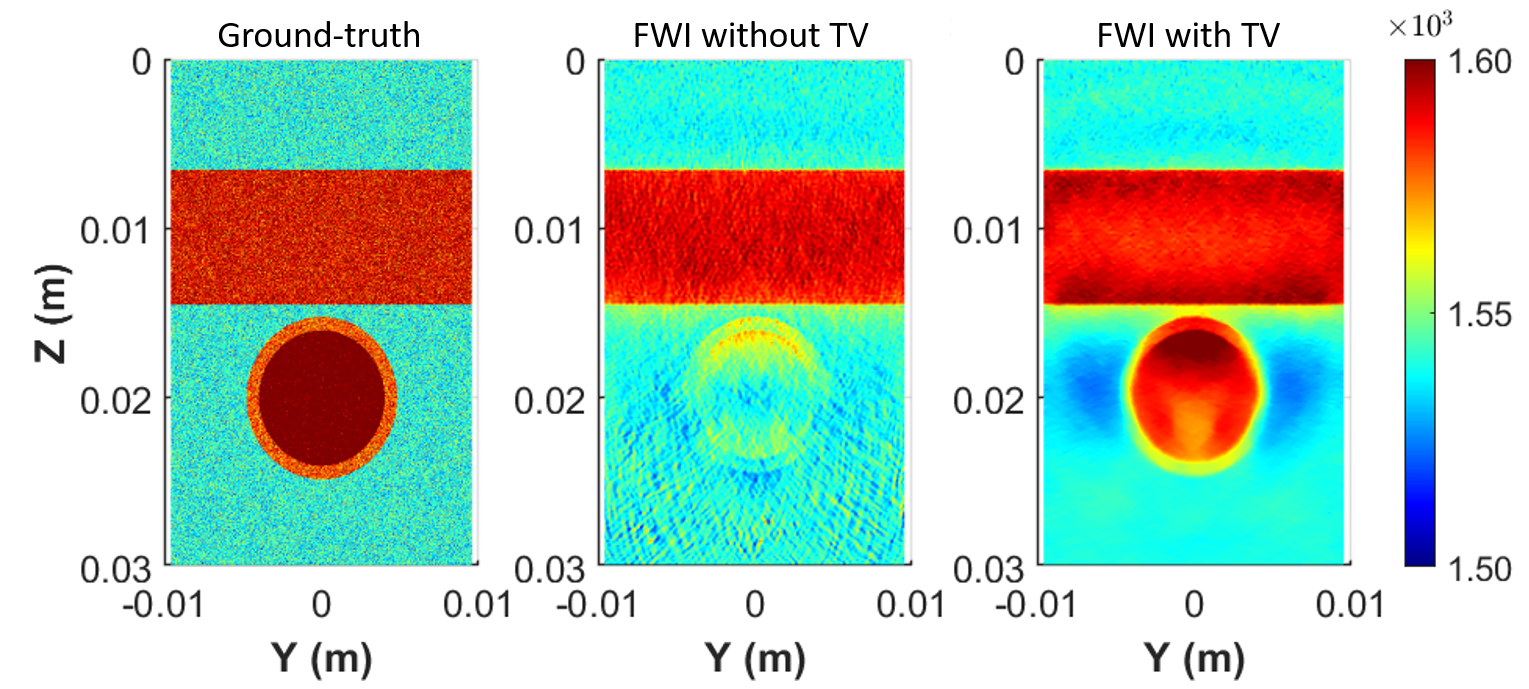}	}
		
		%    \vspace{-0.5cm}/
		\caption{Ground-truth and inverted SoS for the muscle-cyst scenarios.  Z-axis: depth. Left: Ground-truth. Middle: inverted SoS from  FWI without TV. Right: inverted SoS from FWI with TV.  }
		\label{muscle_cyst} 
		%	\vspace{-0.4cm}
	\end{figure}
	
	\textbf{Muscle-cyst scenario:} We further validate our approach in a more challenging scenario where the cyst is shielded by an overlying muscle layer. 
	The muscle layer is modeled with a speed of sound (SoS) of 1590~m/s, an attenuation coefficient of 0.7~dB/cm/MHz, and a thickness of 8~mm. The inversion settings are the same as in the previous experiments, except that the initial model is obtained by blurring the layered background without the target using a Gaussian filter with a standard deviation of 1 mm.
	The ground-truth SoS distribution and the reconstruction results obtained using the proposed approach and the basic FWI method are shown in Fig.~\ref{muscle_cyst}, respectively. 
	The basic FWI fails to recover the internal structure of the high-SoS cyst due to the strong ill-posedness caused by the strong shielding effect of the muscle layer.
	
	\begin{table}[!t]
		\centering
		\caption{Quantitative comparison of SoS reconstruction  in different 2D scenarios. Basic FWI: FWI without TV. FWI-TV: FWI with TV.}
		\label{tab:quantitative_results}
		\begin{tabular}{lcccc}
			\hline
			\multirow{2}{*}{\textbf{Scenario}} 
			& \multicolumn{2}{c}{\textbf{RMSE}} 
			& \multicolumn{2}{c}{\textbf{SSIM}} \\
			\cline{2-5}
			& \textbf{Basic FWI} & \textbf{FWI-TV} 
			& \textbf{Basic FWI} & \textbf{FWI-TV} \\
			\hline
			Simple cyst & 4.3013 & \textbf{3.7857} & 0.6354 & \textbf{0.6386} \\
			Solid cyst  & 24.4291 & \textbf{12.5079} & 0.3311 & \textbf{0.6405} \\
			Muscle-cyst  & 23.7702 & \textbf{13.0278} & 0.4662 & \textbf{0.6119} \\
			\hline
		\end{tabular}
	\end{table}
	Table~\ref{tab:quantitative_results} compares the reconstruction accuracy of basic FWI with TV and the proposed method for the simple,  solid cyst and muscle-cyst scenarios, which is quantified using the root mean square error (RMSE) and the structural similarity index measure (SSIM).  
	%	Let $\mathbf m_{gt}$ denote the ground-truth model and $\hat{\mathbf m}$ the reconstructed result. 
	%	The RMSE is defined as
	%	\begin{equation}
	%		\mathrm{RMSE} = \sqrt{\frac{1}{N_e} \left\lVert \hat{\mathbf m} - \mathbf m_{gt} \right\rVert_2^2}.
	%	\end{equation}
	%	The SSIM between $\hat{\mathbf m}$ and $\mathbf m_{gt}$ is defined as
	%	\begin{equation}
	%		\mathrm{SSIM}(\mathbf m_{gt},\hat{\mathbf m}) =
	%		\frac{(2\mu_m \mu_{\hat{ m}} + C_1)(2\sigma_{m\hat{ m}} + C_2)}
	%		{(\mu_m^2 + \mu_{\hat{ m}}^2 + C_1)(\sigma_m^2 + \sigma_{\hat{ m}}^2 + C_2)},
	%	\end{equation}
	%	where $\mu_m$ and $\mu_{\hat{m}}$ denote the mean values of ${\mathbf m}_{gt}$ and $\hat{\mathbf m}$, 
	%	$\sigma_m^2$ and $\sigma_{\hat{m}}^2$ are the corresponding variances, 
	%	$\sigma_{m\hat{m}}$ is the covariance between ${\mathbf m}_{gt}$ and $\hat{\mathbf m}$, 
	%	and $C_1$ and $C_2$ are small positive constants introduced to stabilize the division. 
	It can be seen that the TV-regularized approach consistently achieves lower RMSE and higher SSIM, with substantial improvements for the solid cyst and muscle-cyst cases.

	\subsection{3D scenario}
	\textbf{Measurement setup}: 
	We now consider a setup that more closely reflects a practical ultrasound acquisition. The transducer is modeled as a finite-length linear array, which introduces out-of-plane wave propagation effects and therefore requires a fully 3D forward and inverse problem. 
	Specifically,we further consider a case when illumination aperture is limited, i.e., the array consists of 64 finite-length line elements with an element spacing of 0.3~mm, resulting in a total lateral aperture of 19.2~mm. Each element has a finite elevation width of 1~cm, consistent with typical linear ultrasound transducers. In the measurement process, each element sequentially transmits while all elements record the backscattered signals.
	To emulate experimentally acquired data, the measurements are generated in the frequency domain at a set of 10 discrete frequencies uniformly distributed between 0.2~MHz and 2~MHz.

\begin{figure}[!t]
		%\centering
		\subfigure[]{	\includegraphics[trim=10 2 10 10, clip, width=1\linewidth]{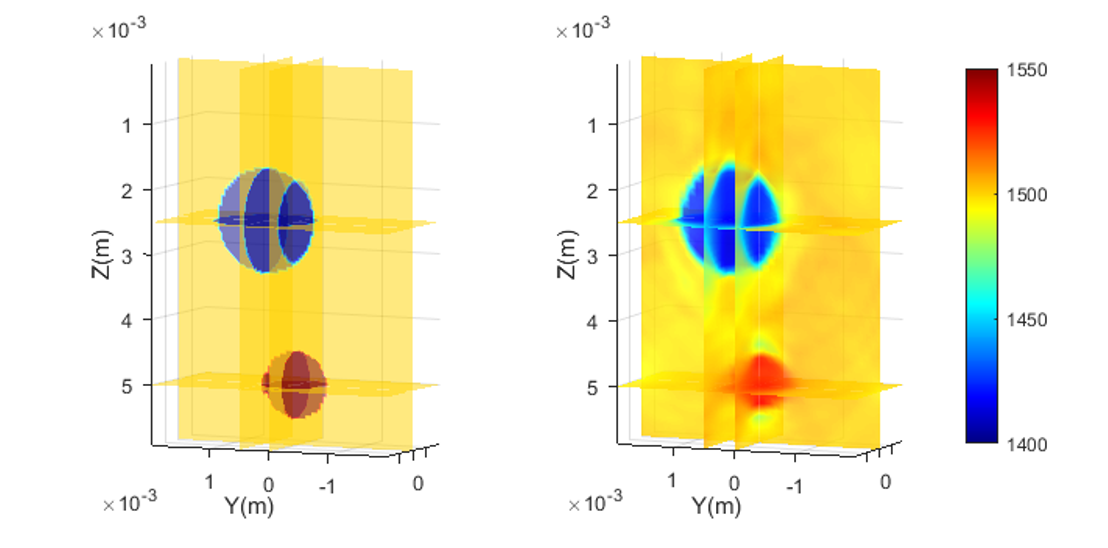}	}
		\subfigure[]{	\includegraphics[trim=10 2 10 10, clip, width=1\linewidth]{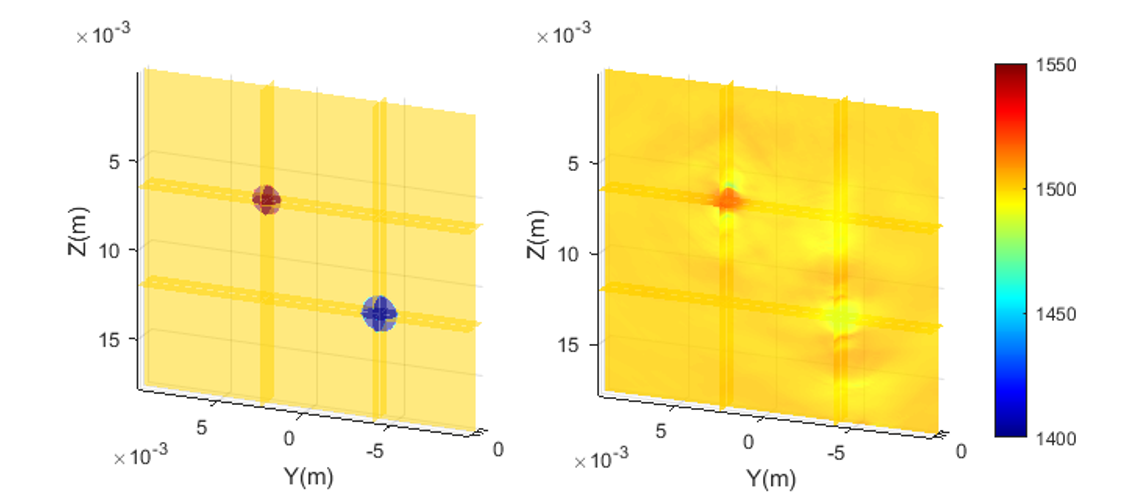}	}
		%    \vspace{-0.5cm}/
		\caption{Ground-truth and inverted SoS in the 3D spherical cyst scenarios via FWI with TV.  Z-axis: depth. (a) Shallow case; (b) Deep case. Left: Ground-truth. Right: inverted SoS.    }
		\label{cyst_3d} 
		%	\vspace{-0.4cm}
	\end{figure}

	\textbf{Test scenario:} 
	Two small spherical cyst with radii of 0.8~mm and 0.5~mm are modeled. 
	One target is assigned a SoS of 1400~m/s and an attenuation coefficient of 0.05~dB/cm/MHz, while the other has a SoS of 1600~m/s and an attenuation of 0.7~dB/cm/MHz. 
	The background  is assumed to be homogeneous with a SoS of 1540~m/s and an attenuation of 0.7~dB/cm/MHz. 
	The two targets are placed at different depths to assess reconstruction performance under depth-dependent sensitivity. 
	The corresponding ground-truth SoS maps are shown in the left panels of Fig.~\ref{cyst_3d} (a) and (b).

	We start the inversion from homogeneous background and set the maximum number of iteration as 20. The reconstructed 3D SoS results of our approach are shown in the right panels of Fig.~\ref{cyst_3d}. 
	Both high- and low-SoS targets can be identified, indicating the feasibility of SoS-based cyst reconstruction for small targets under limited-aperture acquisition. When the targets are located at shallow depths, both the reconstructed shape and internal SoS values are more accurate than those obtained for deeper targets. 
	This performance degradation can be attributed to increased acoustic attenuation in the medium and the reduced amount of scattered energy returning to the receivers. 
	
	We quantitatively compare the reconstructed SoS values within the target spheres  since global metrics alone do not reflect local accuracy in this 3D scenario. In the first case, the reconstructed mean SoS values are 1421.5 m/s and 1517.7 m/s, respectively, preserving the relative contrast between the two targets while exhibiting a bias toward the background.
	In the second case, where the SoS contrast is reversed, the reconstructed mean SoS values are 1512.3 m/s and 1489.6 m/s. Although the overall reconstruction remains visually plausible, the reduced separation between the reconstructed SoS values indicates limited local contrast resolution under this extremely constrained measurement configuration, motivating further investigation into improved acquisition and regularization strategies.
	
	\section{Conclusions and outlooks}\label{SecVI}
	We presented a frequency-domain FWI framework for quantitative SoS reconstruction using single-sided clinical ultrasound arrays. 
	By exploiting the relatively homogeneous background properties of soft biological tissues, the volume integral equations were adopted to enable efficiently simulate pressure wavefields.
	To address the severe ill-posedness under the limited-aperture acquisition, a TV-regularized optimization solved via ADMM was integrated into the inversion framework. 
	Numerical experiments in both 2D and 3D settings demonstrated that the our approach reconstruct  SoS accurately under clinically realistic acquisition constraints.
	
	Several future works will be conducted to extend to practical clinical applications. First, a data calibration strategy needs to be developed to characterize the incident wavefields of commercial ultrasound transducers \cite{wu2023ultrasound}. Second, the optimized frequency sampling and transmission/reception schemes are needed to enhance signal-to-noise ratio and measurement sensitivity to targets \cite{krebs2009fast}. Third, experimental validation using physical phantoms and in vivo data will be conducted. Finally, to achieve real-time clinical feasibility and improve the quantitative recovery of acoustic attenuation, we will explore accelerated solvers, neural network-based surrogate models and data-driven prior learning approaches designed to regularize the inversion process.
	
	%We present a posterior sampling approach for microwave brain imaging that effectively addresses the limitations of existing methods by leveraging a latent diffusion model as an expressive prior. By computing the expectation over the posterior distribution via sampling, our method achieves high-resolution and faithful reconstructions of head electrical properties. Validated on simulated microwave datasets, our approach achieves the smallest model misfit (0.067 \textit{vs.}  $\geq $ 0.085 ) and highest SSIM (0.936 \textit{vs.} $\leq $0.906), and can characterize the imaging uncertainty,  highlighting its reliability for stroke imaging.
	%
	%\begin{table}[!t]
	%	\caption{The mean values of the data misfit, model misfit and SSIM of different methods for the test dataset. \textbf{Bold}: best. }
	%	\centering
	%	\begin{tabular}{|c|c|c|c|c|}
	%		\hline
	%	    Method 	& Occam & TV & GMR & \textbf{Latent diffusion} \\ \hline
	%		Data misfit  & 0.094  & \textbf{0.060}  & 0.068  & 0.065 \\ \hline
	%		Model misfit & 0.112  & 0.111  & 0.085  & \textbf{0.067}  \\ \hline
	%		SSIM  & 0.784  & 0.793  & 0.906  & \textbf{0.936}  \\ \hline
	%	\end{tabular}
	%	\label{tab:simple_table}\vspace{-0.3cm}
	%\end{table}
	%\vspace{-0.3cm}
	\bibliographystyle{IEEEtran}
	\bibliography{IEEEexample_v2}
\end{document}